\documentclass[aps,prl,twocolumn,superscriptaddress,showpacs]{revtex4}

\usepackage{graphicx}
\usepackage{amssymb}

\newcommand{\www}{\langle 111\rangle}
\newcommand{\wwo}{\langle 110\rangle}

\begin{document}

\title{Strongly Non-Arrhenius Self-Interstitial Diffusion in Vanadium}

\author{Luis A. Zepeda-Ruiz}\email[corresponding author: ] {zepedaruiz1@llnl.gov}
\affiliation{Princeton Institute for the Science and Technology of Materials (PRISM),
 Princeton University, Princeton, NJ 08544}
\affiliation{Chemistry and Materials Science Directorate, Lawrence
Livermore National Laboratory, P.O. Box 808, L-371, Livermore, CA
94550} 
\author{J\"{o}rg Rottler}
\affiliation{Princeton Institute for the Science and Technology of Materials (PRISM),
 Princeton University, Princeton, NJ 08544}
\author{Seungwu Han\footnote{current address: Department of Physics, Ewha Womans University, Seoul 120-750, Korea.}}
\affiliation{Princeton Institute for the Science and Technology of Materials (PRISM),
 Princeton University, Princeton, NJ 08544}
\author{Graeme J. Ackland}
\affiliation{School of Physics, University of Edinburgh, Edinburgh EH9 3JZ, United Kingdom}
\author{Roberto Car}
\affiliation{Princeton Institute for the Science and Technology of Materials (PRISM),
 Princeton University, Princeton, NJ 08544}
\author{David J. Srolovitz}
\affiliation{Princeton Institute for the Science and Technology of Materials (PRISM),
 Princeton University, Princeton, NJ 08544}

\date{\today}

\begin{abstract}
We study diffusion of self-interstitial atoms (SIAs) in vanadium via
molecular dynamics simulations.  The $\www$-split interstitials are
observed to diffuse one-dimensionally at low temperature, but rotate
into other $\www$ directions as the temperature is increased.  The SIA
diffusion is highly non-Arrhenius.  At $T<600$ K, this behavior arises
from temperature-dependent correlations.  At $T>600$ K, the Arrhenius
expression for thermally activated diffusion breaks down when the
migration barriers become small compared to the thermal energy.  This
leads to Arrhenius diffusion kinetics at low $T$ and diffusivity
proportional to temperature at high $T$.
\end{abstract}

\maketitle 

The creation and migration of self-interstitial atoms (SIAs) are
critical for microstructural evolution of materials in a variety of
situations, such as in the high energy radiation environment of
nuclear reactors~\cite{young2} and in ion
implantation~\cite{nordlund}.  Although SIA formation energies are
much larger than typical thermal energies, they form in abundance
during collision cascades induced by impinging energetic particles.
SIAs in metals are typically very mobile (i.e., their migration
barriers are relatively small) and hence play an important role in
controlling the rates of many microstructural processes in such
applications, in particular the phenomenon of void swelling.

Since SIA properties and mobilities are very difficult to determine
experimentally, one often employs computer simulations
~\cite{delarubia91,wirth97,soneda98,wirth99}. For example,
simulations of body centered cubic (bcc) iron (and several other bcc
metals), have shown that SIAs preferentially lie along $\wwo$
orientations but rotate into $\www$-directions, where they can migrate
easily using the crowdion configuration as transition state. Other
simulations have suggested that SIA migration in vanadium is very
similar to that in
Fe~\cite{minashin96,morishita97,morishita00,morishita01}.  However,
these empirical interatomic potential-based simulations are not
consistent with recent first principles calculations that clearly show
that the lowest energy SIA configuration in V is a $\www$-split
interstitial, rather than the $\wwo$-split configuration found in
Fe~\cite{han02}.  Interestingly, the first principles calculations
also revealed that the $\www$-oriented SIA migration energy is
extraordinarily small ($\le0.01$eV), which explains the experimental
observation of diffusion down to 4 K \cite{coltman75}.  SIA migration
in Fe and V must therefore differ in their microscopic mechanisms.

We perform a series of molecular dynamics (MD) simulation of SIA
migration in V using an improved interatomic potential for
V~\cite{potential} (refit to experimental and first principles data
~\cite{han02} to reproduce the stable interstitial configuration) to
address this discrepancy.  In particular, we examine SIA diffusion as
a function of temperature to determine the SIA migration mechanisms.
We find that while SIA migration in V is similar to that in bcc Fe in
many respects, its temperature dependence is highly unusual,
exhibiting strongly non-Arrhenius behavior and correlation effects.
It is this non-Arrhenius behavior that is the focus of the present
Letter.

The SIA is introduced in the form of a stable $\www$-split
interstitial and equilibrated for 10 ps at fixed temperature using a
Langevin thermostat. The simulation was then switched to a
microcanonical (NVE) ensemble in order to study SIA migration
dynamics. Simulations were run at temperatures between 100 and 2000 K
in a cubic simulation cell of edge length $10\,a_0$, where $a_0$ is
the temperature dependent-lattice parameter (the linear thermal
expansion coefficient was $8.4\times 10^{-6}$ K$^{-1}$).  The
diffusivity was measured by averaging over several 1 ns
simulations. The interstitial position was identified by dividing the
space into Wigner-Seitz (W-S) cells centered around each perfect
crystal lattice site. Interstitials are located in W-S cells
containing more than one atom.

\begin{figure}[t]
\includegraphics[width=8cm]{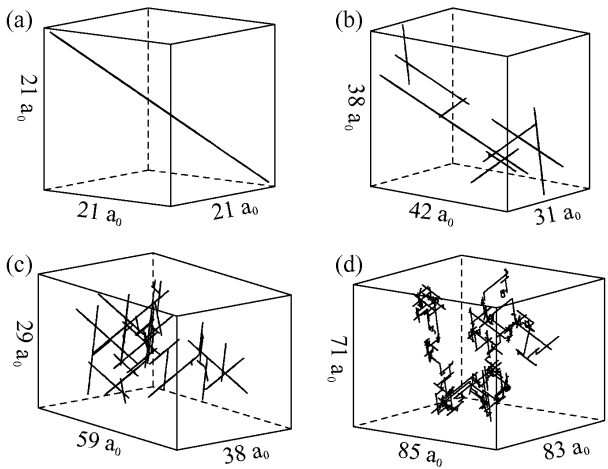}
\caption{\label{traj}Typical trajectories of migrating SIAs for
temperatures of (a) 300 K, (b) 700 K, (c) 900 K and (d) 1400 K.}
\end{figure}

Representative trajectories of the $\www$-split interstitial center of
mass, collected over the whole 1 ns simulations, are shown in
Fig.~\ref{traj}.  For each temperature, more than 1000 jumps were
observed, where an SIA jump is the exchange of an atom between
neighboring W-S cells.  As seen in Fig.~\ref{traj}, the interstitial
migration mechanism is strongly temperature dependent. For low and
intermediate temperatures (100-600 K) the $\www$-split interstitial
executes a fully one-dimensional (1D) random walk along a
$\www$-direction during the $1$ ns simulation, as shown in
Fig.~\ref{traj}(a).  At $T \sim$ 700 K, the $\www$-split interstitial
begins to make infrequent rotations from one $\www$- to another
$\www$-direction.  This results in a three-dimensional (3D) trajectory
that consists of long 1D random walk segments with abrupt
reorientations, as seen in Fig.~\ref{traj}(b). As the temperature
increases, the frequency of the rotation events increases and the
lengths of the 1D trajectory segments decrease. At high temperatures,
the rotation events become very frequent, leading to nearly isotropic
diffusion (Fig.~\ref{traj}(d)).

Although these trajectories appear to be qualitatitively similar to
those reported for other bcc metals (i.e. Fe and
Mo)~\cite{wirth97,osetsky97,pasianot00,marian01}
(cf. Fig.~\ref{traj}(d) and Fig. 5 in~\cite{pasianot00}), they differ
in the elementary migration mechanism.  The stable interstitial is the
$\www$-split configuration in V, but the $\wwo$-split configuration in
Fe and Mo. In the Fe and Mo cases, the split interstitial sits in the
$\wwo$-orientation until it is thermally activated into one of the
$\www$-directions where it can migrate easily before returning to a
$\wwo$-orientation~\cite{pasianot00}.  There are no relaxation events
of this type in interstitial migration in V.  Here, the stable
$\www$-split interstitial migrates long distances and only requires
significant thermal activation to reorient or rotate.

\begin{figure}[t]
\includegraphics[width=8cm]{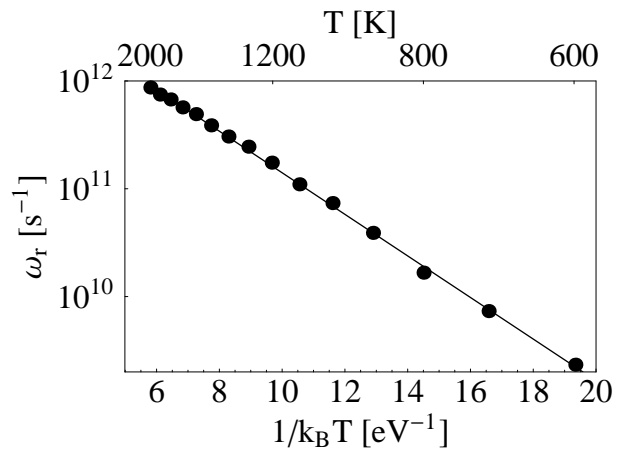}
\caption{\label{rot-fig}Frequency of rotations as a function of
temperature. The solid line is an Arrhenius fit to the data with slope
$\Delta E_r =0.44$ eV, preexponential factor $\nu_0=1.3\times
10^{13}s^{-1}$.}
\end{figure}

The temperature dependence of the rate of rotation of the split
interstitials from one $\www$ direction to another, $\omega_r$, in V
is shown in Fig.~\ref{rot-fig}. The data is well described by a
conventional Arrhenius fit of the form $\omega_r=\nu_0\exp{[-\Delta
E_r/k_BT]}$, suggesting that rotation is a simple thermally activated
process.  The activation energy, $\Delta E_r=0.44$ eV, is consistent
with the energy difference between the $\www$ and $\wwo$
configurations computed in first-principles (0.35 eV) and static
calculations using the new interatomic potential (0.4 eV) \cite{potential}.

\begin{figure}[t]
\includegraphics[width=8cm]{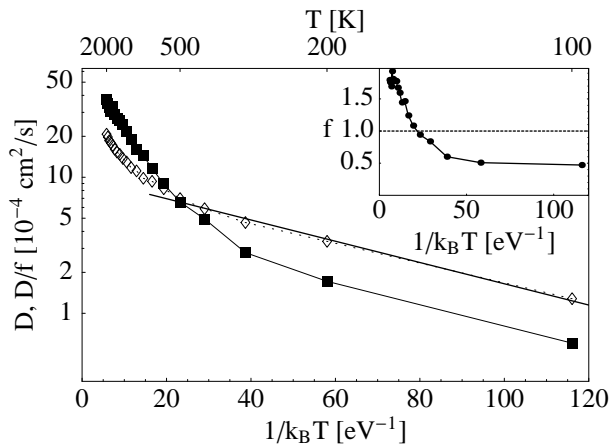}
\caption{\label{diffusivity}A plot of the diffusivity of the
$\www$-split interstitial in the form suggested by the Arrhenius
relation. The filled symbols $(\blacksquare)$ correspond to the
measured diffusivity $D$ and open symbols $(\diamond)$ to $D$
normalized by the correlation factor $f$ (see text).  The straight
line is a low temperature fit to the $D/f$ data, corresponding to
$\Delta E_d=0.018$ eV and a pre-exponential factor of
$\nu_0=1.5\times10^{12}s^{-1}$ in the Arrhenius form. Statistical
errors are of order the symbol size. The inset to the figure shows the
variation of $f$ as a function of 1/$k_BT$.}
\end{figure}

The diffusivity $D$ (solid symbols) of the $\www$-split interstitial
is shown in Fig.~\ref{diffusivity} for a temperature range between
100 K and 2000 K. $D$ was determined from $D=\langle R^{2}(t)\rangle/2t$
for 1D diffusion, where the mean squared displacement $\langle
R^{2}(t)\rangle$ was calculated from the trajectory using standard
averaging procedures~\cite{guinan77}. Note that although the rotations
at higher $T$ change the dimensionality of the diffusion path, they do
not contribute to transport and hence the diffusion mechanism is
always one-dimensional.  If the diffusivity were Arrhenius
($D/a_0^2=\nu_0 \exp{[-\Delta E_d/k_BT]}$), the data in
Fig.~\ref{diffusivity} would lie along a straight line. This is
clearly not the case; Fig.~\ref{diffusivity} shows pronounced
curvature - especially at high temperature. Although Arrhenius
behavior is widely expected for diffusion in the solid state, it is
clearly inapplicable here.

Non-Arrhenius behavior can have several different origins. The energy
barrier could simply change with temperature as a result of thermal
expansion, as has been argued for the self-diffusion in bcc metals via
a vacancy mechanism \cite{eftaxias89,neumann90}.  The magnitude of the
observed deviations from Arrhenius behavior is too large to attribute
to such a mechanism.  The existence of multiple reaction pathways with
different energy barriers can also lead to curvature in
Fig.~\ref{diffusivity}.  However, detailed examination of the atomic
configuration during diffusion shows that there is no change in
diffusion mechanism over the entire temperature range.  Although
rotations are first observed at $\sim 700$ K within the 1 ns duration of the
simulations, strong deviations from Arrhenius behavior are observed
already at lower temperatures.  A third possibility is that the degree
of correlations in the diffusion process (i.e., particle jumps retain
memory and the random walk is non-Markovian) is temperature
dependent. Indeed, examination of the SIA trajectories show that the
$\www$-interstitial has a higher probability of jumping back in the
direction from whence it came, rather than forward along the same
direction. At high temperatures, by contrast, this effect appears to
be reversed.

We quantified this observation by measuring a correlation factor for
split interstitial diffusion $f$, defined as $f=2D/D_{b}$, where
$D_{b}$ is the ``bare'' diffusion constant defined as
$D_{b}=l_0^{2}n$. Here, $n$ is the mean number of jumps/second and
$l_0=\sqrt{3}a_{0}/2$ is the jump length (nearest neighbor distance in
the bcc lattice). If the interstitial trajectory is described by a
sequence of jump vectors $\vec l_i$, the correlation factor $f$ is
alternatively given by $f=1+2\sum_i^{n-1}\langle \vec l_i\cdot \vec
l_0\rangle/l_0^2$, i.e. $f=1$ for an uncorrelated random walk. The
inset to Fig.~\ref{diffusivity} shows the variation of $f$ with
1/$k_BT$. At low $T$ ($T\le600$ K), SIA motion is anticorrelated
$(f<1)$.  The effect of the correlations on the diffusivity can be
isolated by plotting $D/f$ rather than $D$ in Fig.~\ref{diffusivity}
(open symbols). The low temperature $D/f$ data $(T\le 500K)$ lie along
a nearly straight line with slope $\Delta E_d=0.018$ eV. Hence, the
temperature dependent correlation factor explains the relatively weak
deviations from Arrhenius behavior at low $T$. The unusually small
migration energy is consistent with experimental observations
\cite{coltman75} and the first principle estimate \cite{han02}.  One
possible origin of the anticorrelations is that the finite relaxation
time of the local environment around the SIA becomes less important as
the thermal energy of the system increases.

\begin{figure}[t]
\includegraphics[width=8cm]{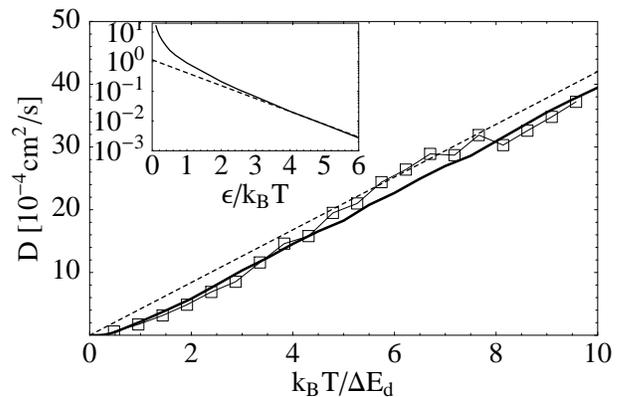}
\caption{\label{lindiff-fig}The diffusivity $D$ $(\blacksquare)$ from
Fig.~\ref{diffusivity} as a linear function of $T$, normalized by
$\Delta E_d$. The thick solid line represents the diffusivity of a
particle in a sinusoidal potential at finite temperature as described
by Eq.~(\ref{model}) for a value of $\gamma=0.1\tau^{-1}$ where
$\tau=\sqrt{m\sigma^2/\epsilon}$.  The dashed line is the free
particle limit of the same model, $D=k_BT/\epsilon m\gamma$. The inset
shows an Arrhenius plot of the diffusivity of the particle in the
sinusoidal potential as a function of $\epsilon/k_BT$ as a solid line
and the straight dashed line has slope one.}
\end{figure}

As the temperature is increased beyond 300 K, the correlation factor
rises quickly to a value greater than unity.  $f>1$ is very unusual
and may be thought of as correlated interstitial hopping over several
barriers without completely thermalizing in between.  This
interpretation is consistent with the fact that the correlation
corrected diffusivity, $D/f$, only yields a straight line at low $T$
but not at high $T$. We note that where $D/f$ is rising quickly, the
energy barrier $\Delta E_d$ obtained in the low temperature regime
(where the behavior is Arrhenius) is smaller than the thermal energy.
This is the source of the multiple interstitial hops at high $T$.
Conventional derivations of activated escape over barriers
\cite{haenggi90} usually assume $\Delta E_d\gg k_BT$ .

Rather than trying to apply the Arrhenius description to interstitial
self-diffusion in this system, a more general model is the motion of a
particle in a periodic potential at all temperatures (from
$k_BT\ll\Delta E_D$ to $k_BT\gg\Delta E_D$).  In the limit that the
barrier height is completely negligible relative to the energy of the
heat bath (i.e., a free particle), standard arguments predict
$D=k_BT/m\gamma,$
where $m$ is the mass of the particle and $\gamma$ a relaxation time
scale associated with a velocity-dependent friction force.  Since this
free particle diffusion coefficient is expected to be a linear
function of $T$, we replot the data from Fig.~\ref{diffusivity} on a
linear temperature scale (Fig.~\ref{lindiff-fig}). In this representation,
the data is nearly linear, albeit with weak curvature at low
temperature.  This suggests that SIA diffusion in V is free particle
like at high temperature ($D\sim T$), but follows the normal
Arrhenius, hopping dynamics at low $T$ ($D\sim e^{-\Delta E_D/{k_BT}}$).
The deviation from linearity at low temperature and the deviation from
Arrhenius behavior at high temperature suggests that a cross-over is
occuring between the free and hopping particle limits.

In order to better understand this transition, we explicitly consider a
simple one-dimensional particle of mass $m$ diffusing in a sinusoidal
potential by numerically solving the Langevin equation
\begin{equation}
\label{model}
m \ddot x-\gamma \dot x=\epsilon/2\cos{[x/\sigma]}+\eta,
\end{equation}
where $\eta$ is a Brownian white noise. Inserting values of $m$ and
$\sigma$ for V and $\epsilon=\Delta E_d$, this model yields nearly
linear diffusivity for $1<k_BT/\epsilon<10$ for reasonable values of
$\gamma$, followed by a crossover into Arrhenius behavior (see inset
of Fig.~\ref{lindiff-fig}) for $k_BT/\epsilon\lesssim 1$.  Changing
the value of $\gamma$ shifts the temperature at which the transition
from the particle hopping to free particle behavior is observed. This
excellent agreement between the MD results and model predictions
demonstrates that the observed strongly non-Arrhenius interstitial
diffusion in V is a direct consequence of the relative magnitudes of
the activation energy and the thermal energy.
  
It is interesting to compare the situation described here for V to
that for bcc Fe.  The crowdion mechanism enabling the easy
interstitial migration along $\www$ directions in V is also available
in Fe, and estimates for the migration barrier $\Delta E_d=0.04$ eV
\cite{wirth97} are similar to the V case. However, measurements of the
apparent activation energy for diffusivity, analogous to the one
presented here (albeit over a smaller temperature range between 700 K
and 1200 K) yield much larger values of $\Delta E_d=0.12$ eV
\cite{marian01} or $\Delta E_d=0.17$ eV \cite{soneda01}. This larger
effective barrier is due to the fact that the $\www$-split
interstitial must be thermally excited from the $\wwo$-state, i.e. the
easy-diffusion configuration is not populated at all times as in
V. The fraction of time during which the $\www$-split interstitial is
available for transport is given by $F=P_{\www}/(P_{\www}+P_{\wwo})$,
where $P_{\www}\sim\exp{[-\Delta E_f/k_BT]}$ and
$P_{\wwo}\sim\exp{[-\Delta E_b/k_BT]}$ are probabilities for the
interstitial to transform from the $\wwo$ state to the $\www$ state
and back, respectively. Therefore, the interstitial diffusivity in Fe
is the product of the diffusivity that the interstitial would have if
it was always in the $\www$ orientation (like in V) and $F$.  For
$\Delta E_f\gg\Delta E_b$, the effective activation energy for
interstitial diffusivity in Fe is $\sim(\Delta E_D+\Delta E_f)$, but
for $\Delta E_b\gg\Delta E_f$, it is $\sim\Delta E_D$.  Clearly, this
implies for Fe that $\Delta E_f\gg\Delta E_b$.

Molecular dynamics simulations of self-interstitial diffusion in bcc V
were performed over an unusually wide temperature range (100-2000
K). Interstitial atoms in the $\www$-split configuration migrate very
fast one-dimensionally along $\www$ directions during the 1 ns
simulations.  As $T$ is increased above $600$ K, rotations of the
split-interstitial from one $\www$ orientation to another occur with
increasing regularity.  The rotations can be described by Arrhenius
kinetics with activation energy $\Delta E_r=0.44 eV$. At temperatures
$T<600$ K, the diffusion exhibits significant anticorrelations.  An
Arrhenius analysis of the data (corrected for these anticorrelations)
yields a very small migration energy barrier $\Delta E_d=0.018$ eV.
For $T>600K$, $\Delta E_d$ is much smaller than the thermal energy and
the Arrhenius expression is no longer applicable. The diffusivity then
crosses over from Arrhenius to free particle type diffusion with
increasing $T$. The fact that this type of diffusion is a linear
function of $T$ rather than Arrhenius, as usually assumed, could have
important implications for predicting the lifetimes of reactor
components in vanadium and other bcc metals.

\begin{acknowledgments}
We thank A.~F.~Voter, G.~H.~Gilmer, and J.~A.~Caro for useful
discussions.  This work was performed under the auspices of the
U.~S.~Department of Energy, Office of Fusion Enery Sciences
(DE-FG02-01ER54628) and Lawrence Livermore National Laboratory under
Contract No. W-7405-Eng-48.
\end{acknowledgments}


\end{document}